\documentclass[prl,twocolumn,aps,showpacs]{revtex4}
\begin{document}
\title{Entanglement of Formation for Gaussian States}
\author{J. Solomon Ivan}
\email{solomon@imsc.res.in}
\affiliation{The Institute of Mathematical Sciences,  CIT Campus, Taramani, 
Chennai 600 113, India}
\author{R.  Simon}
\email{simon@imsc.res.in} 
\affiliation{The Institute of Mathematical Sciences,  CIT Campus, Taramani, 
Chennai 600 113, India}

\date{August 11, 2008} 

\begin{abstract}
The entanglement of formation (EOF) is computed for 
arbitrary two-mode Gaussian states. Apart from a conjecture, our 
analysis rests on two main ingredients.  
 The first is  a four-parameter canonical form  we develop for 
the 
covariance matrix, one of these parameters acting as a measure of EOF,   
 and the second is a  
generalisation of the EPR correlation, used in the work of Giedke {\em 
et al} [Phys. 
Rev. Lett. {\bf 91}, 107901 (2003)], to noncommuting variables. The 
conjecture itself is in respect of an extremal property of this 
generalized EPR correlation.
\end{abstract}
\maketitle

Entanglement is an essential resource for many quantum information processing 
tasks, and hence it is important to be able to quantify this resource.
A reasonable set of 
demands lead, in the case of bipartite pure states, to a simple and unique measure 
for this resource: it is the von Neumann entropy  of either 
subsystem\,\cite{ent1,ent2,ent3}. For mixed states, however, many different  
entanglement 
measures continue to be under consideration\,\cite{ent4}. One of these measures 
with an attractive physical motivation is the entanglement of formation 
(EOF)\,\cite{eof}. The asymptotic version of EOF is the entanglement 
cost\,\cite{eof,cost}. EOF is defined as an infimum: 
\begin{eqnarray*}
{\rm EOF}\,(\rho) \equiv {\rm inf}\,\{\,\sum_j p_j E(\psi_j)\, \,|\,\, 
\rho = \sum_j p_j |\psi_j\rangle\langle\psi_j|\,\}\,.\nonumber 
\end{eqnarray*}
The infimum is to be taken over all realizations of 
the given mixed state $\rho$ as convex sum of pure states, and  
$E(\psi_j) \equiv  
S({\rm tr}_B[|\psi_j\rangle\langle\psi_j|])$  where $S(\cdot)$ is the 
von Neumann 
entropy. EOF has been 
computed in closed form for {\em arbitrary} two-qubit 
states\,\cite{eof-qubits}, and  
 for highly 
symmetric states like the isotropic states\,\cite{eof-iso} and the Werner 
states\,\cite{eof-werner}.

Gaussian states, whose entanglement originates in  
nonclassicality of the squeezing type\,\cite{dutta},   
have played a distinguished role in quantum information in respect 
of continuous variable systems\,\cite{cv-qi}. Their use in 
 teleportation\,\cite{tele1,tele2} and quantum 
cryptography\,\cite{crypt} has been 
demonstrated.
 Questions related to their separability\,\cite{sep1,sep2,sep3,sep4} and 
distillability\,\cite{distil} have been resolved. More recently, 
analytic 
expression 
for their EOF has been obtained in the {\em symmetric} case\,\cite{eof-gaussian1}. 
  This notable achievement seems to be the first computation 
of EOF  
for states of infinite rank. These authors exploit a certain 
{\em extremality}  
that the two-mode-squeezed vacuum enjoys in respect of the 
Einstein-Podolsky-Rosen (EPR) correlation\,\cite{epr}   on the one hand  and 
entanglement on the other.  Further analysis of EOF in this case 
has been made\,\cite{marian} from the 
viewpoint of Bures distance

An interesting Gaussian-state-specific generalisation of EOF, the 
{\em Gaussian entanglement of formation}, has also been 
explored\,\cite{eof-gaussian2,illuminati}.  
 But the EOF of  asymmetric Gaussian states  has remained an open 
problem\,\cite{open}.

In this Letter we compute, under a conjecture, the EOF for arbitrary 
two-mode Gaussian states.  Our 
analysis rests on two  principal ingredients. The first one is a  
four-parameter canonical 
form we develop for the covariance matrix; one of these parameters proves to be a 
measure of EOF. The second one is  a family of  generalised EPR correlations 
for {\em noncommuting} pairs of nonlocal variables; this 
family is indexed by a continuous 
parameter $\theta$. And the conjecture is in respect of an extremal 
property of this generalised EPR correlation.

\noindent
{\em Canonical Form for Covariance Matrix}\,:
Given a two-mode Gaussian state, with the mode on Alice's side described by
canonical quadrature variables $x_{A},\,p_{A}$ and that on  Bob's side
by $x_{B},\,p_{B}$, we can assume without loss of generality that the first
moments of all four variables vanish\,\cite{sep2,eof-gaussian1}. 
 Such a zero-mean Gaussian state  is   fully  described
 by the covariance matrix\,\cite{sep2,eof-gaussian1}
\begin{equation}
V_{G}= \frac{1}{2}
\left[ \begin{array}{cccc}
\alpha \beta n & 0 & \beta k_{x} & 0 \\
0 & {\alpha}^{-1} {\beta}^{-1} n & 0 & -{\beta}^{-1} k_{p} \\
\beta k_{x} & 0 & {\alpha}^{-1}\beta m & 0 \\
0 & -{\beta}^{-1} k_{p} & 0 & \alpha {\beta}^{-1} m \\  
\end{array} \right]\,,
\end{equation}
where the phase space variables are assumed to be arranged in the order
$(x_{A},p_{A},x_{B},p_{B})$ $\equiv$ $\xi$, and we have retained through the
parameters $\alpha$, $\beta$ $>$ $0$ the freedom of independent local
 unitary (i.e., symplectic) scalings on the $A$ and $B$ sides. This 
freedom will be used shortly.

Note that $V_{G}$ is left with no correlation  between the `spatial' 
variables
$x_{A},\, x_{B}$ and the `momentum' variables $p_{A},\, p_{B}$. Thus it is
 sometimes convenient to view $V_{G}$ as the direct sum of $2 \times 2$ matrices:
\begin{eqnarray*}
& V_{G} = X_{G} \oplus P_{G}\,, & \nonumber \\
& X_{G} = \frac{\beta}{2}\, 
\left[ \begin{array}{cc}
\alpha n & k_{x} \\
k_{x}  & {\alpha}^{-1} m \\
\end{array} \right],\,\,\,
P_{G} = \frac{{\beta}^{-1}}{2}\,
\left[ \begin{array}{cc}
{\alpha}^{-1} n &  -k_{p} \\
- k_{p} & \alpha m \\
\end{array} \right].&
\end{eqnarray*}\nonumber 

Let $|\Psi_r\rangle$ denote the standard two-mode-squeezed vacuum state with
squeeze parameter $r$. It takes  the Schmidt form
in the standard Fock basis:
\begin{eqnarray}
| {\Psi}_{r} \rangle & = & \sum_{n=0}^{\infty} c_{n}| n {\rangle}_{A} \otimes
| n {\rangle}_{B} \equiv \sum_{n=0}^{\infty} c_{n}| n, n \rangle\,,\nonumber \\
 c_{n} & = & {\tanh}^{n}r /\cosh r\,.    
\end{eqnarray}
Denoting by $E_r$ the entanglement of $|{\Psi}_{r} \rangle$, 
we have
\begin{eqnarray}
E_{r} = {\cosh}^{2} r \log _2 ({\cosh}^{2} r) -
{\sinh}^{2} r \log _2 ({\sinh}^{2} r)\,.
\end{eqnarray}
The covariance matrix of $|\Psi_r\rangle$ has the form 
\begin{eqnarray}
&& V_{\Psi_r} = X_{\Psi_r} \oplus P_{\Psi_r}\,, \nonumber \\
 X_{\Psi_r} &=&
\frac{1}{2}\,  
\left[ \begin{array}{cc}
C & S \\
S  & C 
\end{array} \right], \,\,\,
P_{\Psi_r} = \frac{1}{2}\,  
\left[ \begin{array}{cc}
 C & - S \\
- S & C 
\end{array} \right], \nonumber\\
 C& \equiv&\cosh 2r,~~ S \equiv \sinh 2r\,. 
\end{eqnarray} 

\noindent
{\em Proposition} 1\,: Given a two-mode covariance matrix $V_{G}$, the 
local 
scale
parameters $\alpha$, $\beta$ can be so chosen that $V_{G}$ gets recast in the
form
\begin{eqnarray*}
&& V_0 = \frac{1}{2}\,  
\left[ \begin{array}{cccc}
C + u\,c^2 & 0 & S+ u\,cs  & 0 \\
0 & C + v\,c^2& 0 & - S -v\,cs \\
S + u\, cs & 0 & C+ u\,s^2 & 0 \\
0 & - S - v\,cs & 0 & C + v\,s^2 
\end{array} \right],\nonumber\\
&& C \equiv \cosh 2r_0,\; S\equiv\sinh 2r_0;~~c \equiv \cos \theta_0,\; 
s \equiv \sin \theta_0\,.\nonumber  
\end{eqnarray*}
{\em Note:} We will call $V_0$ {\em the canonical form} of a two-mode 
covariance matrix;
our results below will justify this elevated status. 
We assume without loss of generality $\,n \ge m \,$ or, 
equivalently, $0 < \theta_0 \le \pi/4.\,$   For a given 
$V_{G}$ there will be two solutions for the above form. Canonical form will
 always refer to the one with the smaller squeeze parameter $r_0$, 
which is ensured 
by the restriction 
\begin{eqnarray}
\tan \theta_0 \ge \tanh r_0\,.
\end{eqnarray}
 This condition proves central to our analysis. Its origin may be 
 appreciated by inverse two-mode-squeezing the Gaussian 
state $V_0$ until it becomes just 
separable, and noting that there exists a range of further squeezing in 
which the {\em mixed} Gaussian state remains separable before becoming 
inseparable again. 
 The parameters $u,\,v\ge 0$. {\em The essence of the canonical 
form is that $V_0$ differs from the covariance matrix of a 
two-mode-squeezed vacuum $|\Psi_{r_0}\rangle$ by a positive matrix which 
is a direct  sum of 
two singular $2\times 2$ matrices which are, modulo signature of the 
off-diagonal elements, multiples of one another}. 

\noindent 
{\em Proof}\,: The canonical form demands, as a necessary condition, 
that $\alpha$, $\beta$, and $r$ be chosen to meet  
\begin{eqnarray}
{\rm det}(X_{G} - X_{\Psi_r}) = 0\,, &\,\,\,\,{\rm det}(P_{G} - P_{\Psi_r}) = 
0\,. 
\end{eqnarray}
These being two constraints on three parameters, one will expect to get a
one-parameter family of solutions to these constraints. For each such solution
we may denote the vector annihilated by the singular matrix 
$X_{G} - X_{\Psi_r}$ by $(\sin \theta,\,-\cos \theta)$, and that 
annihilated by
$P_{G} - P_{\Psi_r}$ by $(\sin {\theta}^{'},\,\cos {\theta}^{'})$. 
 The canonical form corresponds to that solution 
 for which  ${\theta}^{'} = \theta$; it is this degenerate value that 
equals 
$\theta_0$ of the canonical form.
 
That there exists such a degenerate value can be seen as follows. 
 We may fix the scale parameter $\alpha$
through $\alpha = \sqrt{{m}/{n}}$, and then solve Eqs.\,(6) for $\beta$
and $r$, the smaller $r$ being the relevant one. We will find 
$\theta = {\pi}/{4}$ and ${\theta}^{'} < {\pi}/{4}$ in this case.
On the other hand if we take $\alpha = \sqrt{{n}/{m}}$ and then solve
Eqs.\,(6), we will find ${\theta}^{'} = {\pi}/{4}$ and $\theta 
<{\pi}/{4}$.
It follows from continuity that there exists an intermediate value 
${\alpha}_{0}$
for the parameter $\alpha$, in the range $\sqrt{{m}/{n}}
< {\alpha} <\sqrt{{n}/{m}}$, for which ${\theta}^{'} = \theta\,
\,(< \pi/4$ since $n>m)$. And this  yields the canonical form.

Viewed alternatively, the canonical form $V_0$ places the following two 
requirements on
the scale factors $\alpha$, $\beta$:
\begin{eqnarray}
&&\frac{{\rm det}X_{G} -1/4}{{\rm det}P_{G} -1/4} 
=  \frac{{\rm tr}({\sigma}_{3} X_{G})}{{\rm tr}({\sigma}_{3} P_{G})}\,,
\nonumber\\
&&{\rm det}(X_{G} - {\sigma}_{3} P_{G}{\sigma}_{3}) = 0\,,
\end{eqnarray}
where ${\sigma}_{3}$ is the diagonal Pauli matrix. These are simultaneous
equations in $\alpha$, $\beta$, and solving these equations yields, in 
terms of $n,\,m,\,k_x,\,k_p$,  the
 values of $\alpha,\,\,\beta$  corresponding to the canonical form.

Two special cases may be noted. 
 If $m=n$
we have $\alpha =1\,$(since $\sqrt{n/m} =\sqrt{m/n}$), and hence $\beta 
= 
\sqrt{(n - 
k_{p})/(n 
- 
k_{x})}$,
so that the canonical squeeze parameter $r_0$ is given by 
$e^{-2r_0} = \sqrt{(n - k_{x})(n - k_{p})}$, reproducing the results of
Ref.\,\cite{eof-gaussian1}. The parameter $\theta_0$ always equals 
${\pi}/{4}$ in 
this (symmetric) case.
 On the other hand,  if $k_{x} = k_{p} = k$, 
the canonical form corresponds to 
$\alpha = \beta =1$,  and one obtains $r_0$ by simply solving  
\begin{equation}
{\rm det}
\left[ \begin{array}{cc}
n-{\rm cosh}2r_0 & k - {\rm sinh}2r_0 \\
 k - {\rm sinh}2r_0 & m-{\rm cosh}2r_0
\end{array} \right] = 0\,,
\end{equation}  
which yields this closed-form expression for $r_0$:
\begin{eqnarray*}
\cosh (2\eta - 2r_0) &=& \frac{nm -k^2 +1}{\sqrt{(n+m)^2 - 
4k^2)\,}},\nonumber \\
{\rm e}^{\,\pm2\eta} &\equiv& \frac{(n+m)\, \pm\, 2k}{\sqrt{(m+n)^2 
\,-\,4k^2\,}}.\nonumber
\end{eqnarray*}

\noindent
{\em Generalised EPR Correlation}\,:
To proceed further, we need to generalise the familiar EPR 
correlation\,\cite{eof-gaussian1}.
Given any bipartite state $|\psi\rangle$, define
\begin{eqnarray}
x_{\theta} &=& \sin \theta\, x_{A} - \cos \theta \,x_{B},~ p_{\theta}
= \sin \theta\, p_{A} + \cos \theta\, p_{B} \,,\nonumber\\
&&~~~{\Lambda}_{\theta}(\psi) = 
\langle\psi|(x_{\theta})^2|\psi\rangle +
\langle\psi|(p_{\theta})^2|\psi\rangle\,. 
\end{eqnarray}
In defining ${\Lambda}_{\theta}(\psi)$ 
we have assumed $\langle\psi|x_{\theta}|\psi\rangle \,=\,0\,=\,  
\langle\psi|p_{\theta}|\psi\rangle$; if this is not the case  
then $x_\theta$ and $p_\theta$ in ${\Lambda}_{\theta}(\psi)$ 
 should be replaced by $x_\theta - \langle\psi|x_{\theta}|\psi\rangle$ 
and  $p_\theta - \langle\psi|p_{\theta}|\psi\rangle$ respectively.  
 Clearly, the usual EPR correlation\,\cite{eof-gaussian1} corresponds 
to $\theta = {\pi}/{4}$. While $x_{{\pi}/{4}},\, 
p_{{\pi}/{4}}$
commute, the generalised EPR (nonlocal) variables 
$x_{\theta},\,p_{\theta}$ 
{\em do not commute}, 
and hence the name generalised EPR correlation  
 for ${\Lambda}_{\theta}({\Psi})$;  
 indeed, we have $[x_\theta,\,p_\theta] = -i\cos 2\theta$.
  For the two-mode-squeezed vacuum $| {\Psi}_{r} \rangle$ 
   the  generalised
EPR correlation reads 
\begin{equation}
{\Lambda}_{\theta}({\Psi}_{r}) = \cosh 2r - \sin 2 \theta \sinh 2r\,.
\end{equation}

Let us combine the quadrature variables of the oscillators of Alice 
and Bob into 
boson operators $a = {(x_{A} + i p_{A})/\sqrt{2}}$ and 
$b = {(x_{B} + i p_{B})/\sqrt{2}}$. Then, 
 ${\Lambda}_{\theta}(\psi)$ has this 
expression quadratic in the boson variables:
\begin{eqnarray}
{\Lambda}_{\theta}(\psi) & =& \langle 
\psi|\hat{\Lambda}_\theta|\psi\rangle,\nonumber\\
\hat{\Lambda}_{\theta}& =& 1 + 2{\sin}^2 \theta\,
 a^{\dagger} a + 2{\cos}^2 \theta \, b^{\dagger} b  \nonumber \\ 
&&~~~~~ - 2\cos \theta \sin \theta(a b +  a^{\dagger} b^{\dagger})\,.  
\end{eqnarray}
We may call $\hat{\Lambda}_{\theta}$ the {\em generalised EPR operator}.

The entanglement of $|\Psi_r\rangle$ monotonically increases with 
increasing 
value of the squeezing parameter $r$. In order that 
$\Lambda_\theta(\Psi_r)$ 
 be useful as an entanglement measure  of $|\Psi_r\rangle$ it should,
 for fixed value of $\theta$, decrease with increasing $r$. The restriction
 $\tan \theta \ge \tanh r $, 
encountered  earlier in Eq.\,(5) from 
a different perspective, simply ensures this.  
Through the monotonic relationship (3) between $r$ and $E_{r}$, we will  
view  this
constraint as a restriction on the allowed range of values of $\theta$, for
a fixed value of entanglement.

Given a squeezed state 
$|\Psi_r\rangle$, let us denote by 
$|\Psi^{\prime}_r\rangle$ the state obtained 
from $|\Psi_r\rangle$  by independent local canonical 
transformations\,\cite{sep2} 
$S_A,\,S_B\in Sp(2,R)$, acting respectively on the 
oscillators of Alice and Bob.

\noindent
{\em Proposition} 2\,: 
 We have $\Lambda_\theta(\Psi^{\prime}_r)
\ge \Lambda_\theta(\Psi_r)$,  
 $\forall \,\theta$ in the range $1\ge \tan \theta \ge \tanh r$  
and for all $S_A,\,S_B \in Sp(2,R)$.

\noindent 
{\em Proof}\,:  Clearly, 
$\Lambda_\theta(\Psi^{\prime}_r) = \frac{1}{2}\,\{\,\cosh 2r [\,\sin ^2 
\theta \,{\rm 
tr} (S_AS_A^T) + \cos ^2 \theta \,{\rm tr} (S_B S_B^T)\,] 
 - \sin 2\theta \sinh 2r \,{\rm tr}\,(\sigma_3S_A\sigma_3S_B^T)\,\}$. 
If $e ^{\pm 
\gamma_A}$ are the singular values of $S_A$, and $e^ {\pm \gamma_B}$ 
those of $S_B$, then 
${\rm tr} (S_AS_A^T)= 2\cosh 2\gamma_A$,  
${\rm tr} (S_BS_B^T)= 2\cosh 2\gamma_B$, and  
  ${\rm tr} (\sigma_3S_A\sigma_3S_B^T) \le 2\cosh (\gamma_A +\gamma_B)$.   
Thus the difference $\Delta(\gamma_A,\,\gamma_B) \equiv 
 \Lambda_\theta(\Psi^{\prime}_r)
 - \Lambda_\theta(\Psi_r)$ obeys 
$\Delta(\gamma_A,\,\gamma_B) \ge 
  \cosh 2r [\,\sin ^2 \theta (\cosh 2\gamma_A -1) 
+ \cos ^2 \theta (\cosh 2\gamma_B - 1)\,] 
 - \sin 2\theta \sinh 2r [\,\cosh (\gamma_A +\gamma_B)-1\,]$. 
 It is easily seen that $\Delta(\gamma_A,\,\gamma_B)$ is extremal 
at $\gamma_A=\gamma_B= 0$ corresponding to the standard squeezed state 
$|\Psi_{r}\rangle$. To show that this extremum is indeed minimum we 
  note that the determinant of the Hessian 
matrix of the right hand side, evaluated at $\gamma_A = 0 = \gamma_B$, 
is 
proportional to $\sin 2\theta \cosh 2r - \sinh 2r$,  and hence is 
positive if and only if $\tan \theta \ge \tanh r$. 

Once again we see a role for the requirement 
$\tan \theta \ge \tanh r$. 
Let the equivalence $V_G \sim V_0$ denote the 
fact that the corresponding  Gaussian states 
 are connected by a local canonical transformation.  
 The fact that $M\equiv V_0 -V_{\Psi_{r_0}}\ge 0$ implies  
$\Lambda_{\theta_0}(\rho_{V_0})\ge 
\Lambda_{\theta_0}(\Psi_{r_0})$. In view of Proposition~2     
 this implies 
 $\Lambda_{\theta_0}(\rho_{V_G})\ge 
\Lambda_{\theta_0}(\rho_{V_0})\ge 
\Lambda_{\theta_0}(\Psi_{r_0})=  
\cosh 2r_0 - \sin 2\theta \,\sinh 2r_0$   
for any  Gaussian state 
$V_G$ connected to $V_0$ by local canonical transformation. 
This assigns an  alternative  meaning to the canonical 
parameter  $r_0$:

\noindent
{\em Proposition} 3\,: Given a Gaussian state described by $V_G 
\sim V_0$,  the canonical squeeze parameter 
$r_0$ is the smallest $r$ for which the matrix 
 inequality $V_G - V_{\Psi^{\prime}_r} \ge0$ is true. 

It is well known that the two-mode-squeezed vacuum has several 
extremal properties of interest to 
entanglement\,\cite{extremal,eof-gaussian1}.  
 It seems that this state enjoys one more such distinction, this time  
in respect of our generalised EPR correlation.

\noindent
{\em Conjecture} 1\,: Among all bipartite states of fixed entanglement 
 numerically equalling $E_{r}$, and
for every $\theta$ in the range 
$\tanh r \le \tan \theta \le 1$, 
the two-mode-squeezed 
vacuum $|{\Psi}_{r}\rangle$ yields the least value for the generalised EPR
correlation ${\Lambda}_{\theta}(\cdot)$. In other words, no state $|\psi \rangle$
with entanglement $E(|\psi \rangle) \leq E_{r}$ can yield a generalised
EPR correlation ${\Lambda}_{\theta}(\psi) <     
{\Lambda}_{\theta}({\Psi}_{r})$, for any $\theta$ in the range $\tan 
\theta \ge \tanh r$

The special case $\theta = {\pi}/{4}$ is the basis of
the important work of Ref.\,\cite{eof-gaussian1}. Hence the present assertion can 
be viewed as
a generalisation of their Proposition~$1$. 
 
The original EPR correlation $\Lambda_{\pi/4}(\cdot)$ continuously 
decreases to zero with increasing entanglement. But this is not true 
of the generalised EPR correlation $\Lambda_{\theta}(\cdot)$.  
 
Let us denote by $r_\theta$ the value of $r$ determined by a given 
value of $\theta$ 
through the equation   
$\tan \theta = \tanh r $, and let $\theta _r$ denote the value of 
$\theta$  so determined by $r$. 
 Then, for a given numerical $E_r$, the relevant 
range for $\theta$ in Conjecture~1 is  $\theta_r \le \theta \le \pi/4$. 

\noindent
{\em Proposition} 4\,: The generalised EPR correlation 
$\Lambda_\theta(\cdot)$ 
obeys the basic inequality 
$\Lambda_\theta(\cdot)\ge \cos 2\theta$. 
The two-mode-squeezed vacuum saturates this inequality if and 
only if the squeeze parameter $r$ solves  $\tanh r = \tan \theta$.

\noindent
{\em Proof}\,: It is clear that the relations  
$\tan \theta = \tanh r $, 
$\sin 2\theta =\tanh 2r$, 
and $\cos 2\theta = (\,\cosh 2r \,)^{-1}$ are equivalent to 
one another, and so also are the inequalities 
$\tan \theta \ge \tanh r $, 
$\sin 2\theta \ge\tanh 2r$, 
and $\cos 2\theta \le (\,\cosh 2r \,)^{-1}$.    
 Now consider the transformation $(a,\,b) 
\to  U(r)(a,\,b)U(r)^{\dagger}$ where   
 $U(r) = \exp\{ \,r 
(a^{\dagger}b^{\dagger} - a b)\,\}$ 
is  the unitary two-mode-squeeze operation:  
\begin{eqnarray*}
a \to a \cosh r - b^{\dagger} \sinh r ,\;\;
b \to b \cosh r - a^{\dagger} \sinh r .\nonumber
\end{eqnarray*}
This implies the following transformation for the anticommutator 
$\{b,b^{\dagger}\} \equiv bb^{\dagger} + b^{\dagger}b$\,:  
\begin{eqnarray*}
\{b,b^{\dagger}\} 
&\to& (\,b^{\dagger}b -  a^{\dagger}a\, )  
 + \frac{1}{2}(\{a,a^{\dagger}\} +  \{b,b^{\dagger}\} )\cosh 
2r\nonumber\\   
 &&~~~~~~~~- (\,ab +  a^{\dagger}b^{\dagger}\,)\sinh 2r   \nonumber\\
 &&=\cosh 2r\,\hat{\Lambda}_{\theta_r},\; \;\theta_r 
\equiv \arctan(\tanh r).\nonumber
\end{eqnarray*}
Since $\{b,b^{\dagger}\} \ge 1$,  so is also its unitary transform 
 $\cosh 2r\,\hat{\Lambda}_{\theta_r}$.  
That is,  $\hat{\Lambda}_{\theta_r} \ge (\,\cosh 2r\,)^{-1} 
=\cos 2\theta_r$.

Thus, saturation of the inequality 
 ${\Lambda}_{\theta_r}(\psi^{\prime}) \ge \cos 2\theta_r$ is equivalent 
to the condition 
$\langle\psi|\{b,b^{\dagger}\}|\psi\rangle =1$, where 
 $|\psi^{\prime}\rangle = U(r)|\psi\rangle$.  
 A pure state which satisfies 
$\langle\psi|\{b,b^{\dagger}\}|\psi\rangle =1$, 
 is of the form $|\psi\rangle = |\phi\rangle_A \otimes 
|0\rangle_B$, 
where $|\phi\rangle_A$ is {\em any}  vector in Alice's Hilbert space 
${\cal H}_A$.
 It follows that states saturating the inequality 
 ${\Lambda}_{\theta_r}(\rho) \ge \cos 2\theta_r$ constitute the set 
$\{\, \rho = U(r)  \rho_A \otimes|0\rangle_B\,_B\langle 0| 
U(r)^{\dagger}\,\}$,  where 
$\rho_A$  is any (pure or mixed) state 
of Alice's oscillator. 
Finally, Conjecture~1 claims that   among all these  
states saturating this inequality 
 the two-mode-squeezed vacuum 
$|\Psi_{r_\theta}\rangle$, corresponding to  the choice
$\rho_A = |0\rangle_A\,_A\langle 0|$,  
has the least entanglement.  

\noindent
{\em Entanglement of Formation}\,:
With the canonical  form and the generalised  EPR correlations in hand, 
we are now fully equipped to  compute the 
 EOF of an arbitrary two-mode Gaussian state.

\noindent
{\em Proposition} 5\,: Given an inseparable zero-mean two-mode Gaussian 
state 
$\rho_{V_0}$  with covariance matrix $V_0$  specified in the canonical form 
by  $u,\,v,\,\theta_0$ and $r_0$  with $u, \,v\ge 0$ and 
$0<\tanh r_0 \le \tan \theta_0 \le 1$, 
its EOF equals $E_{r_0}$, the entanglement of the squeezed 
vacuum $|\Psi_{r_0}\rangle$.

\noindent
{\em Proof}\,: 
The fact that $M\equiv V_0 -V_{\Psi_{r_0}} \ge 0$ guaranties  
 that $\rho_{V_0}$ can be 
realized as a convex sum of displaced versions 
$D(\xi)|\Psi_{r_0}\rangle$ of the squeezed vacuum 
state  $|\Psi_{r_0}\rangle$,  all of which have the same 
entanglement $E_{r_0}$ as $|\Psi_{r_0}\rangle$:
\begin{eqnarray*}
\rho_{V_0} \,\sim\, \int {\rm d} ^2 \xi D(\xi)|\Psi_{r_0}\rangle \langle 
\Psi_{r_0}|D^\dag (\xi) \exp (-\frac{1}{2}\xi^T M^{-1} \xi).\nonumber  
  \end{eqnarray*}
Here $D(\xi)$ is the unitary phase space displacement operator. 
 The rank of  $M$ equals $2$, and both $M^{-1}$  and the  
 two-dimensional integral refer to the restriction of the  phase space 
 variable $\xi$ to the range of $M$. 
 
Since a specific 
ensemble realization with average entanglement $E_{r_0}$ is exhibited, 
EOF$(\rho_{V_0}) \le E_{r_0}$.
  On the other hand, evaluation of the generalised EPR correlation 
  ${\Lambda}_{\theta} (\rho_{V_0}) = 
 {\rm tr}\,(\hat{\Lambda}_{\theta} \rho_{V_0})$,    
for the particular value of $\theta$ occurring in 
$V_0$ shows that    
 $ {\Lambda}_{\theta_0} (\rho_{V_0}) = \cosh 2r_0 - \sin 2\theta_0 \sinh 
2r_0$. 
And by  Conjecture~1, this implies  
EOF$(\rho_{V_0}) \ge E_{r_0}$. We have thus proved EOF$(\rho_{V_0}) =  
E_{r_0}$.

 An attractive feature of the canonical form of the covariance matrix is 
that the two-mode-squeezing $U(r)$ acts on it in a 
covariant or form-preserving manner.   

\noindent
{\em Proposition }6\,: Under the two-mode-squeezing transformation 
$U(r)$ we have 
\begin{eqnarray*}
 V_0(r_0,\theta_0,u,v) &\to&    
V_0(r_0^{\prime},\theta_0^{\prime},u^{\prime},v^{\prime})\,;\nonumber\\ 
r_0^{\prime} = r_0 + r,\;\;&&
\sin 2\theta_0^{\prime} = \frac{\sinh 2r + \cosh 2r \sin 2\theta_0} 
{\cosh 2r + \sin 2\theta_0 \sinh 2r},\nonumber\\ 
 (u^{\prime},\,v^{\prime}) &=& (u,\,v)\times (\cosh 2r + \sin 2\theta_0 
\sinh 2r).\nonumber
\end{eqnarray*}

This is easily verified by direct computation. 
 While the canonical squeeze parameter $r_0$ simply gets
 translated by $r$, the parameters $u$ and $v$ get scaled by a 
{\em  common  factor}.  If we  define 
$r_{\theta_0}, \, r_{\theta^{\prime}_0}$ through   
$ \tan \theta_0 \equiv \tanh r_{\theta_0}$ and  
$ \tan \theta^{\prime}_0 \equiv \tanh r_{\theta^{\prime}_0}$, 
the transformation law for $\theta_0$ takes the form   
of translation: $r_{\theta^{\prime}_0} =  r_{\theta_0} + r$. 

As a  consequence of this covariance, the convex decomposition which 
minimizes the average entanglement goes covariantly  to such a    
decomposition under two-mode-squeezing:  the minimal 
decomposition commutes with squeezing. 
 This implies, in particular, the following simple behaviour of 
EOF under squeezing:  
$E_{r_0} \to E_{r_0 + r}$.  

Finally, the just separable Gaussian states on the 
separable-inseparable boundary, correspond to the canonical form with 
$r_0= 0$\,\cite{sep2}. As was to be expected, the condition (5) places 
no restriction on $\theta_0$ in this case.

\end{document}